# Engineering meter-scale porous media flow experiments for quantitative studies of geological carbon sequestration

*Kristoffer Eikehaug[1], Malin Haugen[1], Olav Folkvord[1], Benyamine Benali[1], Emil Bang Larsen[1], Alina Tinkova[1], Atle Rotevatn[2], Jan Martin Nordbotten[3,4], and Martin A. Fernø[1,4]*

[1] Dept. Physics and Technology, University of Bergen
[2] Dept. Earth Science, University of Bergen
[3] Center for Modeling of Coupled Subsurface Dynamics, Dept. of Mathematics, University of Bergen
[4] Norwegian Research Center, Postboks 22 Nygårdstangen, 5838 Bergen

**Abstract**

This technical note describes the FluidFlower concept, a new laboratory infrastructure for geological carbon storage research. The highly controlled and adjustable system produces a strikingly visual physical ground truth of studied processes for model validation, comparison, and forecasting, including detailed physical studies of the behavior and storage mechanisms of carbon dioxide and its derivative forms in relevant geological settings for subsurface carbon storage. The design, instrumentation, structural aspects and methodology are described. Furthermore, we share engineering insights on construction, operation, fluid considerations, and fluid resetting in the porous media. The new infrastructure enables researchers to study variability between repeated $CO_2$ injections, making the FluidFlower concept a suitable tool for sensitivity studies on a range of determining carbon storage parameters in varying geological formations.

## 1. Introduction

We aim to develop a laboratory research infrastructure dedicated to geological carbon storage (GCS) with methodology for repeatable, meter-scale experiments with sufficient precision to allow investigation of any isolated parameter. The two-phase flow of $CO_2$-rich gas and water-rich fluids within complex geological structures, combined with the development of density-driven convective mixing, is difficult to adequately resolve by numerical simulation (Nordbotten et al., 2012; Flemisch et al., this issue). As such, there is a strong need for accurate and reproducible experimental data against which numerical simulation tools can be verified. To this end, this technical note details the construction and operation of a laboratory-scale GSC research infrastructure, which we term "FluidFlower" (see **Figure 1**), with the following a set of characteristics:

A) Meter-scale, quasi-2D experimental systems containing unconsolidated sands that can be arranged to replicate realistic geological structures such as domes, pinch-outs and faults.
B) Operational conditions mimicking real GCS operations can be achieved by rate-controlled injection and localized pressure monitoring.
C) Multiphase flow characteristics such as free gas and $CO_2$ dissolution and concomitant density-driven fingers can clearly be identified visually and quantified by image analysis tools (Nordbotten et al, this issue).
D) Studied geometries can be flushed and reset to the initial state, allowing for studying reproducibility of experimental results (Fernø et al, this issue) as well as variations of operational conditions.

State of the art simulators and experimental methodology builds on a century of global oil and gas exploration and production. While this provides a solid scientific and technological foundation, there remains aspects unique to GCS that require further development. $CO_2$ is physically and chemically very different from crude oil and gaseous organic hydrocarbons: $CO_2$ is highly soluble in water, and its derivative dissolved inorganic carbon (DIC) species both acidify and increase the density of the water. The increase to density allows gravitationally induced convective mixing, an essential sequestration mechanism and a topic of much study over the past decades (Pau et al., 2010; Riaz et al., 2006; Elenius et al., 2012; Erfani et al., 2022). The density change depends on total DIC concentration, in turn depending on dissolution rates, reactivity, convection and diffusion rate, and latent or induced flow in the reservoir. The acidification allows a range of reactions to take place in different geochemical environments leading to effects such as mineralization or grain dissolution, potentially altering the physical properties of reservoirs.

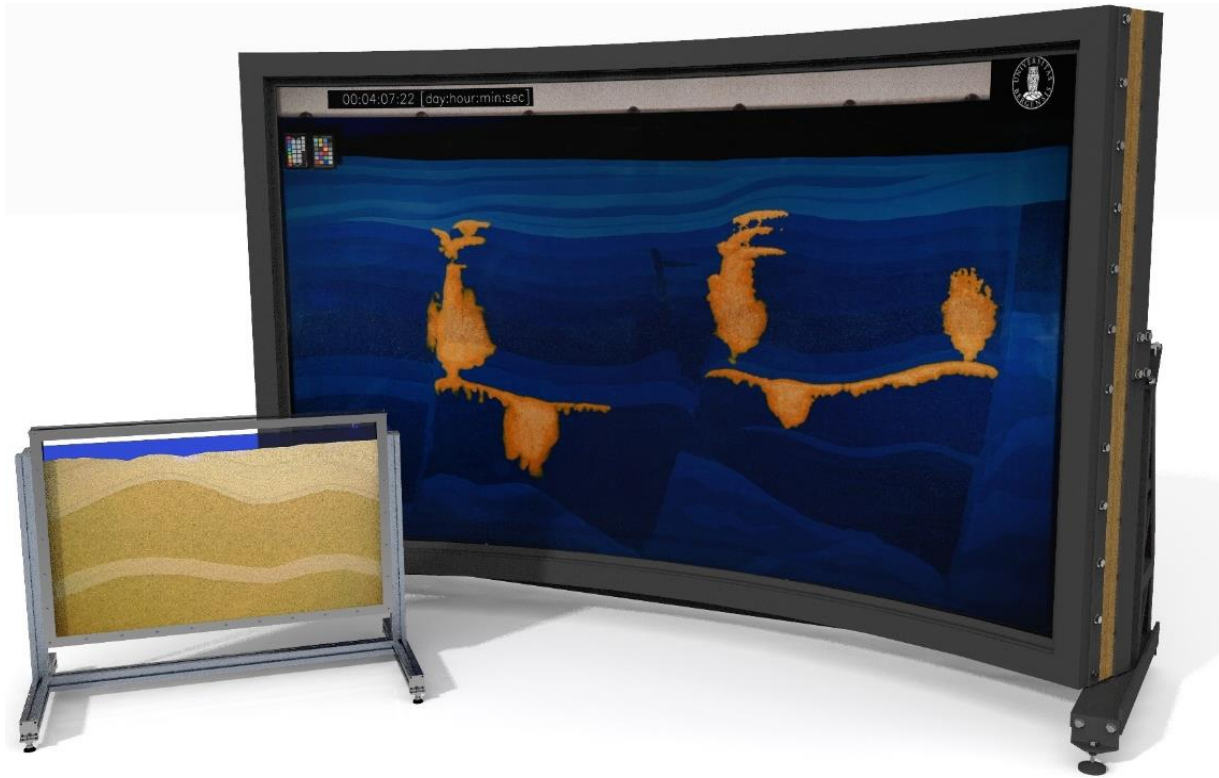

**Figure 1** – FluidFlower flow cell varieties used in the forecasting study. The porous media are built with unconsolidated sands within a vertical, quasi two-dimensional, optically transparent flow cell filled with water. A camera is located on the front side to monitor and record system changes with time-lapsed imaging. Other instrumentation and most operations such as fluid injections occur at the rear side. To achieve the scale of the experiments, operation occurs at standard sea-level pressure and temperature, still preserving the governing porous media physics, relevant displacement processes and trapping mechanisms for subsurface $CO_2$ storage. The FluidFlower concept serves a dual purpose of a research infrastructure for high-quality experimental data, and a vehicle for public outreach and dissemination. The room-scale (right) shown during $CO_2$ injection in a faulted geological geometry motivated by typical North Sea reservoirs. Tabletop-version (left) shown containing an idealized folded geometry. The FluidFlower concept produces quantitative meter-scale experimental data that may be scaled to field-scale (Kovscek et al 2023, this issue) to provide new insight relevant for subsurface GCS. Virtual models for illustration.

The technical note is structured as follows: Chapter 2 presents key features of the FluidFlower concept; Chapter 3 describes the FluidFlower laboratory infrastructure, with technical aspects and considerations for the flow cell; Chapter 4 details experimental operation and capabilities; and Chapter 5 provides the rationale for fluids used. Appendix A details the room-scale flow cell, and its operation and laboratory setup for the $CO_2$ experiments in the 2022 FluidFlower forecasting study (Fernø et al., this issue); Appendix B describes the tabletop flow cells used for methodology development, rapid prototyping and iteration, and smaller meter-scale experiments in (Haugen et al., this issue; Saló-Salgado et al., this issue; Keilegavlen et al., this issue; Eikehaug et al., in press).

## 2. Key FluidFlower features

Here we describe four essential features of the FluidFlower concept that enable meter-scale porous media flow experiments for quantitative studies of geological carbon sequestration.

*Physical repeatability*

Repeated $CO_2$ injection experiments performed on the same porous media geometry is a key capability of the FluidFlower concept. Cycling of fluids allows for investigation of isolated experimental parameters and identification of stochastic elements without the uncertainties and workload associated with rebuilding the geometry for each new experiment. The process of resetting fluids between repeated experiments is designed to keep colors and chemical conditions constant within an experiment series for increased reproducibility (Forecasting study cycling shown in Appendix A and more general example in Appendix B). This technique allows complete tabletop fluid resetting in a few hours and room-scale resetting in a few days.

*Porous media*

The porous media are constructed by depositing unconsolidated material with known properties to quantify observed processes for model verification, comparison, and forecasting. Sand grains should remain within a known and comparatively narrow size distribution, ideally with minimal shape variation to avoid unwanted packing effects. The sands should also be chemically inert unless grain dissolution or similar processes is the intended subject of study. The depositing process is designed to mimic natural underwater sedimentation. The above considerations, the construction of porous media geometries, and the tools used, are further detailed in (Haugen et al., this issue). Settling of unconsolidated sands may be traced using the open-source software DarSIA (Both et al., 2023; Nordbotten et al., this issue).

*Seeing fluid phases*

Differentiating between fluid phases is essential to the viability of the method. Both water and gas are normally transparent and provide little optical contrast to one another. For GCS applications, water and $CO_2$ diffuse into one another, and mixes of two or more fluids are of particular interest. Aqueous concentration of $CO_2$ and DIC is determined by dissolved gas capacity in the water, distance from gas phase $CO_2$, diffusion rate, convective mixing, time, and chemical environment. Dissolution of $CO_2$ into water is an important GCS storage mechanism, and its physical effects in a reservoir has been a topic of interest for some time in the simulation community. To visualize this, we utilized the spontaneous reaction between $CO_2$ and water producing carbonic acid (detailed in chapter 5).

*Data collection*

Time lapse images of the flow cells are captured at intervals synchronized to the experiment time steps, acquired by high-resolution cameras with constant settings to improve reproducibility. Imperfections in the camera optics cause spatial distortion in images, and composite grid images of levelling laser lines provide a reference for correction of such lens aberrations. Images captured (by RGB sensors) never fully represent (full spectrum) true colors, and a standard color palette is included in all images as an image processing reference. Temperature, point and ambient pressure, and all fluid injections are measured and logged (detailed in chapter 4).

## 3. Infrastructure

Flow cells are functionally similar to flattened aquariums, where vertical transparent plates separate observers and instrumentation from sands and fluids. Key flow cell structural considerations are detailed below.

*Flow cell depths*

Flow cell depth is a compromise between observational and operational aspects, limiting sand packing boundary and three-dimensional flow effects. The distance between the flowing gas and viewing window should be small so that diffusion allows for early detection of displacement processes in the third dimension, while allowing sufficient depth for manipulation of sands using sand manipulation tools described in (Haugen et al 2023, this issue). The hydrophilic wall surfaces encourage gas flow within the sands rather than along the walls to minimize the effect of artificial grain structuring along the boundary. For grain sizes typically less than 2 mm associated with unconsolidated sands (e.g., Freeze et al, 1979), a depth on the order of 8-10 times the maximum grain size is preferred to mitigate poor packing conditions and preferential flow along the walls (Chapuis, R.P., 2012).

*Internal forces and hydraulic deformation*

Water exerts significant outward static pressure on the flow cell walls. The hydraulic load scales with the squared height of water in the flow cell, increasing with hydrostatic pressure and surface area (approximately eight tons for room-scale version and ¼ ton for tabletop version). Furthermore, the settling of the unconsolidated sands will lead to additional lateral forces. The large span of the walls argues for the accommodation of measurable and safe elastic deformation, rather than to strive for absolute rigidity to avoid potential brittle failure. Hence, transparent plastics has been the material of choice. The perforations of the rear panel weaken the structural integrity of the plate with unknown localization of material stresses, and a safety margin has been applied to all structural dimensions. The room-scale FluidFlower is curved to further increase rigidity with the front in a state of compression and the rear in a tensile state (see **Figure A1** in Appendix A).

*Viewing window reflections*

The reflective surface of the viewing window causes artefacts in images captured. Images are captured by a camera located in the focal point of a curved flow cell, where any horizontal straight line drawn between the camera and the flow cell viewing window is orthogonal to the latter (see **Figure A5** in Appendix A), allows minimum reflections compared to imaging of a flat flow cell such as the tabletop version. With maximum overlap between direct and reflected camera field of view, lamps and other objects not placed directly between the camera and flow cell remain hidden to the camera. This allows illumination with high incident angle and a compact laboratory footprint. Curvature also increases structural rigidity and allows a larger window span without requiring impractically thick walls to withstand internal forces. Film studio standard high-frequency 'flicker-free' LED lamps are located outside of the camera field of view.

*Construction materials and rear plate perforations*

Materials in contact with the internal volume should have minimum influence on porous media chemistry and flow behavior. The materials must be resistant to the corrosive nature of the salts and fluctuating pH in the system to allow study of the contained system rather than the external. Rear plate perforation limits the instrumentation and tubing required to manipulate and measure the fluids of interest to the rear plate. Hence, the front side remains free of disturbing technical elements. Fluid resetting for repeated experiments occurs through a series of perforations along the lower flow cell boundary where water may be injected to replace the fluids in the system. The rigid flow cell boundary is provided by a double-flange frame for viewing window support. The chassis-like substructures are built to accommodate instrumentation, mobility, and ease of operation. They connect to hinges on the flow cell and are constructed to transfer their load to the chassis bottom where adjustable legs are installed, as well as transport wheels for the room-scale version.

## 4. Experimental operation and capabilities

This chapter provides a general overview of technical instrumentation that enables the injection and quantitative monitoring of $CO_2$ flow, trapping and dissolution in the geological geometries in the flow cells.

*Fluid flow systems*

There are separate control systems for the aqueous and gaseous phase injections (illustrated in **Figure 2**). Water flow is operated with computer-controlled double-piston pumps (e.g., Chandler Engineering Quizix Precision) that connect to the system via gas traps that double as particle traps. Valve manifolds connect gas traps to flow cell ports for controlled injection/production from specific ports, or with a total rate distributed between multiple ports. Manual and/or computer-controlled pneumatic valves may be used. Gas flow is regulated by computer-controlled mass flow controllers (MFC, e.g., Bronkhorst El-Flow Prestige), with gas supplied from pressurized gas canisters. Standard pressure regulators connect the gas bottles to the MFCs via a flow restriction needle valve that reduce pressure fluctuations originating from the spring-loaded pressure regulator mechanism. MFC performance must be tested and tuned prior to all experiments to keep fluctuations within listed instrument uncertainty. Aqueous and gaseous operations should be fully scripted to limit the potential for human error.

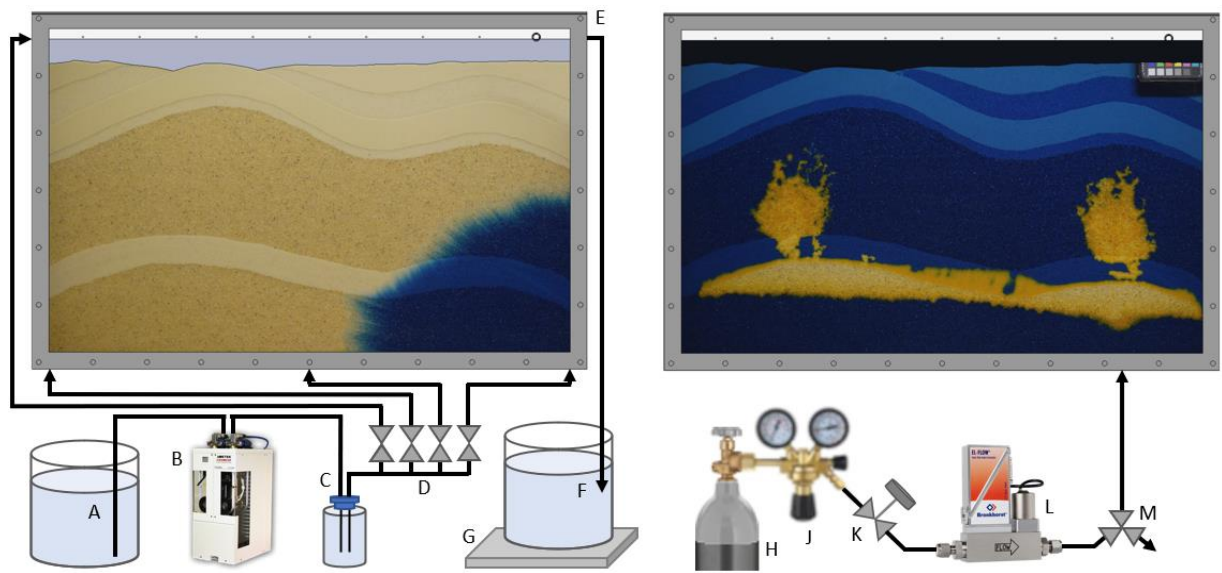

**Figure 2** - Conceptual fluid systems: Water (left) and gas (right). Water from supply (A) to computer-controlled cylinder pump (B), and further through a gas trap (C) before it is directed to the system perforations by valve manifolds (D). Displaced water (overflow) (E) exits an open perforation and is collected by a waste canister (F) sitting on a mass logging scale (G). Gas flows from gas canister (H) with pressure regulator (J), and further through a flow restriction needle valve (K) and a computer-controlled mass flow controller (L) connected to the system perforations or the atmosphere via a three-way valve (M).

*Logging volumetric displacement of water*

A constant water level is maintained due to the passive overflow function of the always-open perforations positioned on the top of the flow cell (cf. **Figure 2**). Water is injected at a constant rate into the free water above the porous media geometry to keep the overflow port water-filled to eliminate surface tension effects. This approach keeps fluctuations in the hydrostatic pressure to a minimum. By logging overflow rates via interval mass measurement, rates of volumetric displacement of water (cf. Appendix A) become detectable and may be coupled to gas phase $CO_2$ volume and its dissolution rate in the system.

*Pressure and temperature*

Pressure transducers connect to chosen perforations for point logging. Typically, fluctuations of interest in the systems occur in the sub-mbar regime and require sensors of adequate precision for any meaningful measurements. Working at sea-level pressure implies atmospheric fluctuations represent significant uncertainty if unattended. Not only may system response disappear in atmospheric noise, but flow experiments using gas phase $CO_2$ are highly dependent on absolute pressure and temperature for both density and dissolution. Temperature is sought to be kept constant during experiment series, yet a gradient has been observed along the height of the flow cells in cold-floor labs. Point logging of temperature is collected by dual-purpose pressure transducers (e.g., ESI Technology GS4200-USB).

*Degassing of the aqueous phase*

Aqueous solutions should be degassed by a vacuum pump (e.g., Edwards RV5) prior to injection to minimize the influence of atmospheric gases dissolved in the water on measured variables. Atmospheric gases in the water affect the $CO_2$ dissolution rates, and Henry's law (see chapter 5) implies that when $CO_2$ dissolves into atmosphere-saturated water, non-negligible quantities of nitrogen and oxygen are expelled from the water. This presents significant challenges for quantitative analysis of time-lapsed image series and resetting of the porous media if unattended. The influence of varying degassing is discussed further in Haugen et al, this issue.

**5. FluidFlower fluids**

Differentiating between fluid phases is essential to the viability of the FluidFlower concept. The $CO_2$ is injected as a dry gas and will partially partition into the formation water (aqueous pH-indicator mix), and this is of particular interest for GCS applications as it is one of the trapping mechanisms which makes the injected $CO_2$ less mobile with time. The equilibrium concentration of dissolved gas in water is proportional to the system pressure and inversely proportional to temperature, and varies between types of gases and combinations thereof. This relation is given by Henry's law as species specific Henry solubility $H^{cp}$ translating to aqueous equilibrium concentration $c_{aq}$ for any type of gas, with $c_{aq}$ proportional to the partial pressure of a gas species.

To visualize dissolution of $CO_2$ into water, we utilized the spontaneous reaction between $CO_2$ and water producing carbonic acid. The acidification allows compounds sensitive to pH changes in the neutral to mildly acidic regime to be an accessible method of aqueous $CO_2$ detection. Common pH indicators are medium-size organic compounds that undergo a configurational change when a proton is added or subtracted from the molecule. The configuration change in turn causes a change in the wavelengths of light absorbed, observed a visible and reversible change of color.

Pure water has a theoretical pH of 7.0, with $[H_3O^+] = [OH^-] = 10^{-7}M$, but with no buffering capacity it is extremely sensitive to impurities. Atmospheric $CO_2$ diffuses into the water and acidification of freshly deionized water can be measured immediately after air exposure. Hence, pure water in equilibrium with atmospheric gasses typically has a pH of approximately 5.8, compared with a pure $CO_2$ atmosphere (emulating conditions inside the flow cells) at approximately pH 4. Several pH indicator compounds in this range exists, typically with transition from a high-absorption color at higher pH towards a lower intensity color (lower pH), complicating precise determination of $CO_2$ concentrations and making the visual appeal mediocre at best. While simple image processing may capture aqueous $CO_2$ contours, the signal strength remains limited.

pH indicators typically have a transition range of ΔpH 1-2, regardless of specific pH transition range. With pH being a logarithmic measure of acid or base concentrations centered around 7, transition ranges closer to 7 distinguish a narrower range of concentrations. We have opted to work under the assumption that widening the pH range of the water phase with a dilute strong base has a measurable yet limited effect on $CO_2$ dissolution and the overall system behavior. This allows more intense contrast colors and distinct transition ranges (cf. **Figure 3**), leading to improved signal strength and the possibility of distinguishing multiple concentration levels of $CO_2$.

Bromothymol blue (BTB, transition pH 6.0-7.6) and methyl red (MRe, transition pH 4.4-6.2) have been used extensively in our experiments. In acidic environment, however, the protonated form of methyl red has a relatively low solubility in water, and high concentrations result in precipitation at pH 6 and below, like that observed in the $CO_2$ injections (Fernø et al, this issue; Haugen et al., this issue; Saló-Salgado et al., this issue).

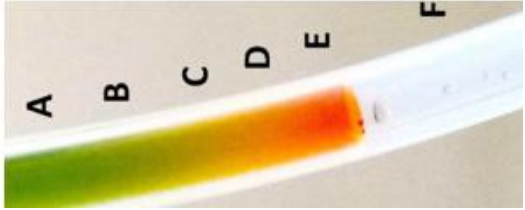

**Figure 3** - Visible $CO_2$ concentration gradient in a mixture of BTB and MRe. A) Solution with no $CO_2$, pH 8. B) BTB transition, pH 7. C) Transition overlap, pH 6. D) MRe transition, pH 5. E) $CO_2$- saturated solution, pH 4. F) gaseous $CO_2$ phase

## 6. Concluding remarks

The FluidFlower concept, as presented in this technical note, represent a new laboratory infrastructure for experimental research to add to the knowledge base for which decisions regarding GCS is made. Details and engineering insights for constructing and operating these highly controlled and adjustable systems are presented for flow cells of different complexity. Both designs have been proven viable and reliable after several experiment series lasting up to a year or more. The combination of these physical flow cells, together with the in-house developed open-source software DarSIA (used to analyze high-resolution time-lapse images), makes it possible to plan and perform a variety of porous media fluid flow experiments on the meter-scale with quantification of key parameters. This provides a unique opportunity to obtain experimental data for validation and calibration of numerical simulation models.

The geological geometries that can be modelled by the FluidFlower are representative of large-scale structures (e.g., large scale faults and folds) and stratigraphic layering (reservoir units, seal units etc.) observed in subsurface reservoir systems, and the FluidFlower concept allows for studying the impact of these geological geometries on $CO_2$ trapping and flow dynamics. Subsurface $CO_2$ trapping mechanisms that can be studied with the FluidFlower concept includes; *structural and stratigraphic trapping* under sealing sand layers; *residual trapping* is seen in regions with intermediate water saturation, and is temporary because of rapid dissolution; *dissolution trapping* is observed almost instantaneously when the injected $CO_2$ dissolves into the water phase; and *convective mixing* which occurs when the denser $CO_2$-saturated water migrates downwards, generating gravitational fingers.

Ultimately, our concept allows observation of spatio-temporal interactions of physical processes of multiphase, multi-component flow during $CO_2$ immobilization in a porous medium at the meter scale with high relevance for GCS applications. We encourage the porous media to explore this experimental method.

**Acknowledgements**
KE and MH are partly funded by Centre for Sustainable Subsurface Resources, Research Council of Norway (RCN) project no. 331841. MH is also funded by RCN project no. 280341. BB is funded from RCN project no. 324688. The authors would like to acknowledge civil engineering interns Ida Louise Mortensen, Mali Ones, Erlend Moen, and Johannes Salomonsen for their laboratory and workshop contributions to the Forecasting study and legacy experiments. Presented tabletop illustrations use images from experiments by Ingebrigt Lilleås Midthjell, Håkon Kvanli, Håkon Stavang, Maylin Elizabeth Ordonez Obando, and Janner Fernando Galarza Alava. Geology input from and room-scale illustration image from geometry developed in collaboration with Robert Leslie Gawthorpe and Casey William Nixon. Roald Langøen, Charles Thevananth Sebastiampilai, Thomas Husebø, Werner Olsen, Thomas Poulianitis, and Rachid Maad have contributed with technical solutions from parts machining to instrument and software prototyping.

**Appendix A – Operations and infrastructure enabling repeated CO$_2$ injections in the forecasting study**

This appendix provides infrastructure details of the room-scale FluidFlower (in-house designation FF2) and operational procedures that supplement information for five repetitive CO$_2$ injections in the 2022 FluidFlower forecasting study (Fernø et al, this issue).

A1 – Dimensions of the room-scale FluidFlower rig

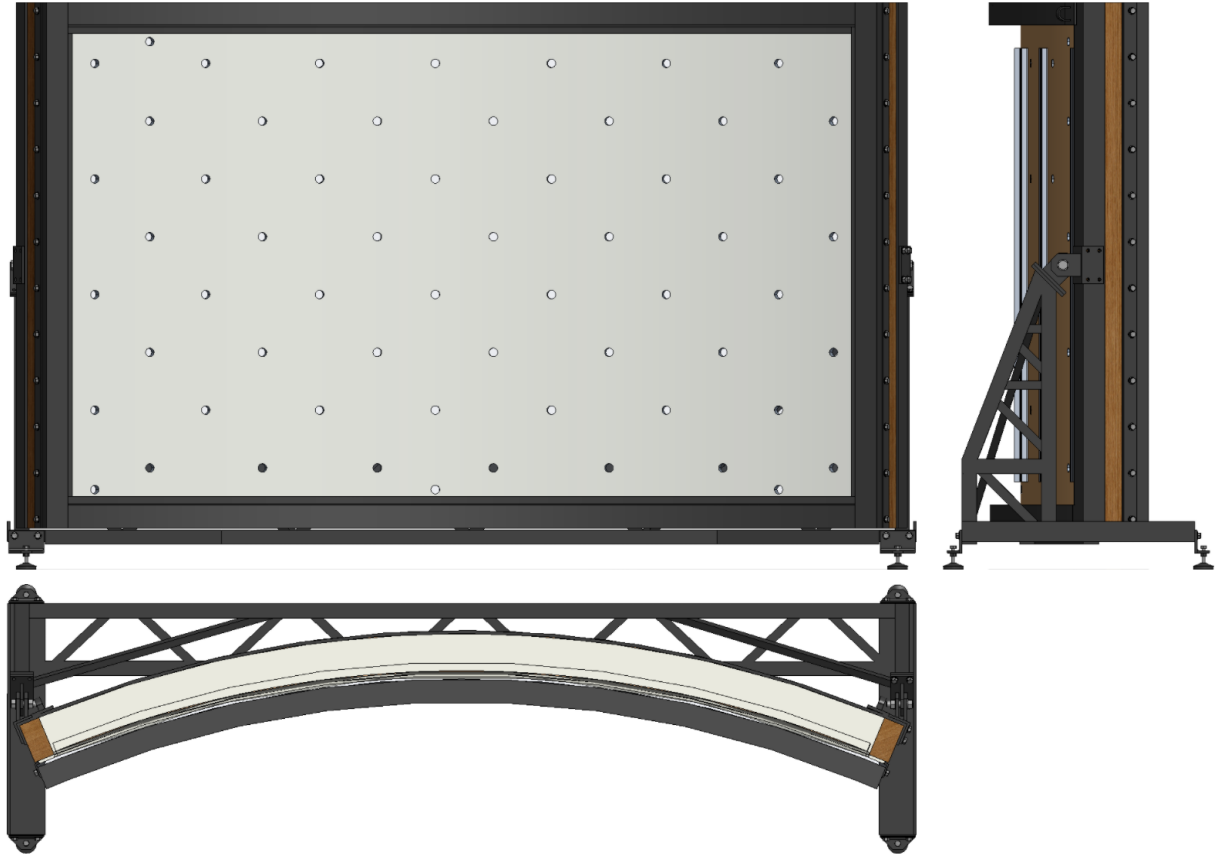

**Figure A1** – Front, side, and top view profiles of the room-scale FluidFlower used for the main CO$_2$ injection experiments in the Forecasting study (Fernø et al, this issue). Instrumentation rails are attached to the rear plate and to the substructure.

**Table A1** – Dimensions for the room-scale rig

| | | |
|---|---|---|
| **Internal measurements** | Height, total [mm] | 1740 |
| | Height of water table [mm] | 1540 |
| | Width:    Inner radius (3600 mm) arc length [mm] | 2860 |
| |             Outer radius (3618 mm) arc length [mm] | 2874 |
| | [a]Depth [mm] (empty) | 18 |
| | Volume (empty) under operational water level [L] | 80 |
| | Volume (empty), total [L] | 90 |
| **Window measurements** | Height, total [mm] | 1600 |
| | Height of visible water table [mm] | 1500 |
| | Width:    Visible radius (3600 mm) arc length [mm] | 2780 |
| | Area, total [m$^2$] | 4.45 |
| **Full structure measurements** | Height [mm] (minimum configuration) | 1990 |
| | Width [mm] | 3175 |
| | Depth [mm] (without legs) | 800 |
| | Mass, empty [kg] (approximate) | 700 |

[a] Only accurate when the flow cells are not under strain (filled with liquids and/or sand)

## A2 – Material choices for the room-scale FluidFlower

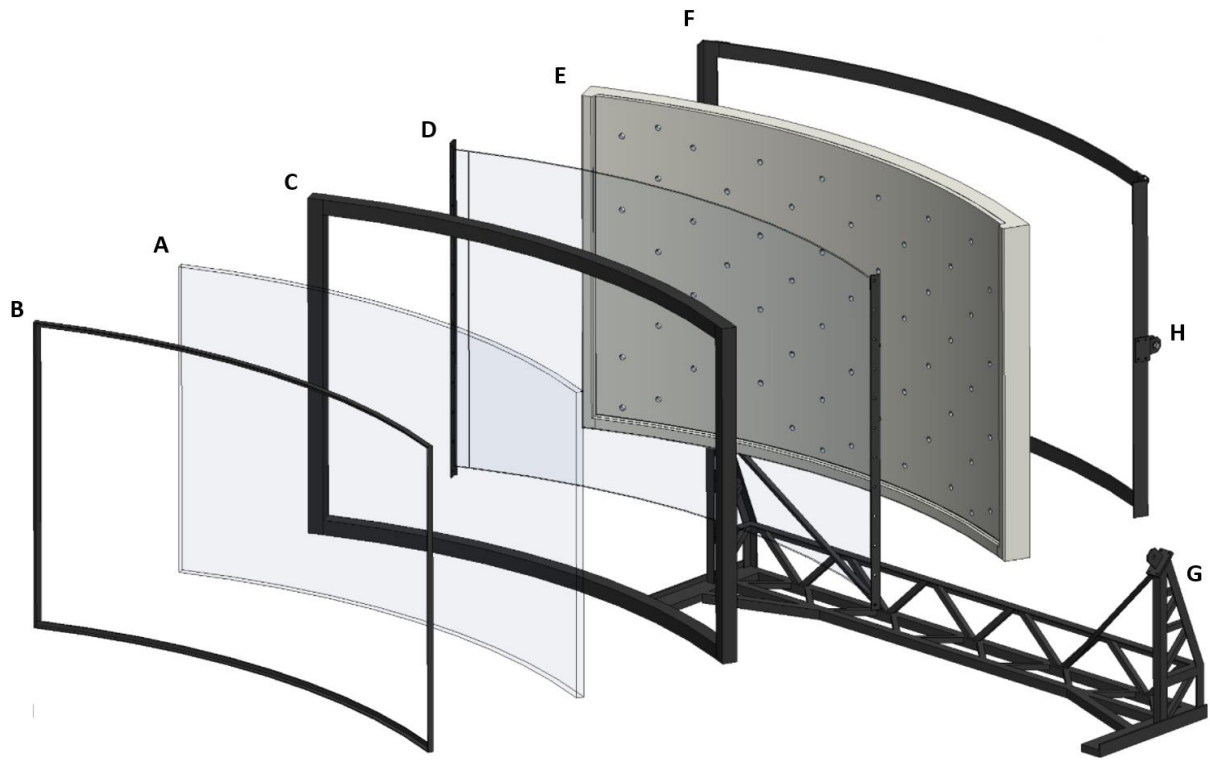

**Figure A2** – Exploded assembly of structural room-scale flow cell components. A) A hot-molded monolithic acrylic window rests on a B) ledge inside the C) front frame and a D) sacrificial polycarbonate plate is clamped to its outer radius. The rigid flow cell boundary is provided by a double-flange steel frame. The front frame is constructed for both conventional flange loading with integrated threads and to resist compression and buckling, as well as to resist the keystone-like lateral force of the acrylic plate arising from deformation under load. The outer radius of the E) rear plate needs only resist stretching, not compression or buckling, and is therefore based on a simpler design where top, side and bottom wooden beams are laminated outward and capped with a heavy F) steel band dimensioned to resist inter-bolt deformation. The G) substructure connects to H) hinges on the rear plate assembly of the flow cell. The substructure is designed using a traditional truss approach and allows some twist to distribute forces between the wheels during transport and to dampen shocks. The front frame assembly is attached to the rear assembly by bolts extending through the circumference of the rear plate from the rear frame.

**Table A2** – Material choices for the room-scale FluidFlower.

|  |  |  |
|---|---|---|
| **Internal materials** | Front wall | Polycarbonate glass |
|  | Rear wall | Epoxy paint |
|  | Bottom/sides | Epoxy paint |
|  | Rear corners | Epoxy paint |
|  | Front corners | Aquarium silicone |
|  | Perforation seal | Aquarium silicone |
|  | Perforations | Stainless steel |
| **External materials** | Structural front wall | Acrylic glass |
|  | Structural rear wall | Epoxy fiberglass Laminated birch plywood |
|  | Flange frames | Mild steel |
|  | Instrumentation rails | Aluminium |
|  | Substructure | Mild steel |

A3 – Perforations in room-scale FluidFlower

The room-scale rig has 56 perforations in a grid layout, plus 3 emergency drains and 1 permanently open overflow drain (cf. **Figure A3**). Perforations are modified cast 316L steel ¾'' marine thru-hull fittings with external tensioning nuts to allow a variety of configurations (see **Figure A4**). The internal flanges rest on polyurethane in a conical indentation in the plate. The internal perforations volume contained either solid or heat-clamped steel tube perforated polyoxymethylene (POM) fittings with their front-facing circumference sealed off with aquarium silicone. A POM cap secures the insert.

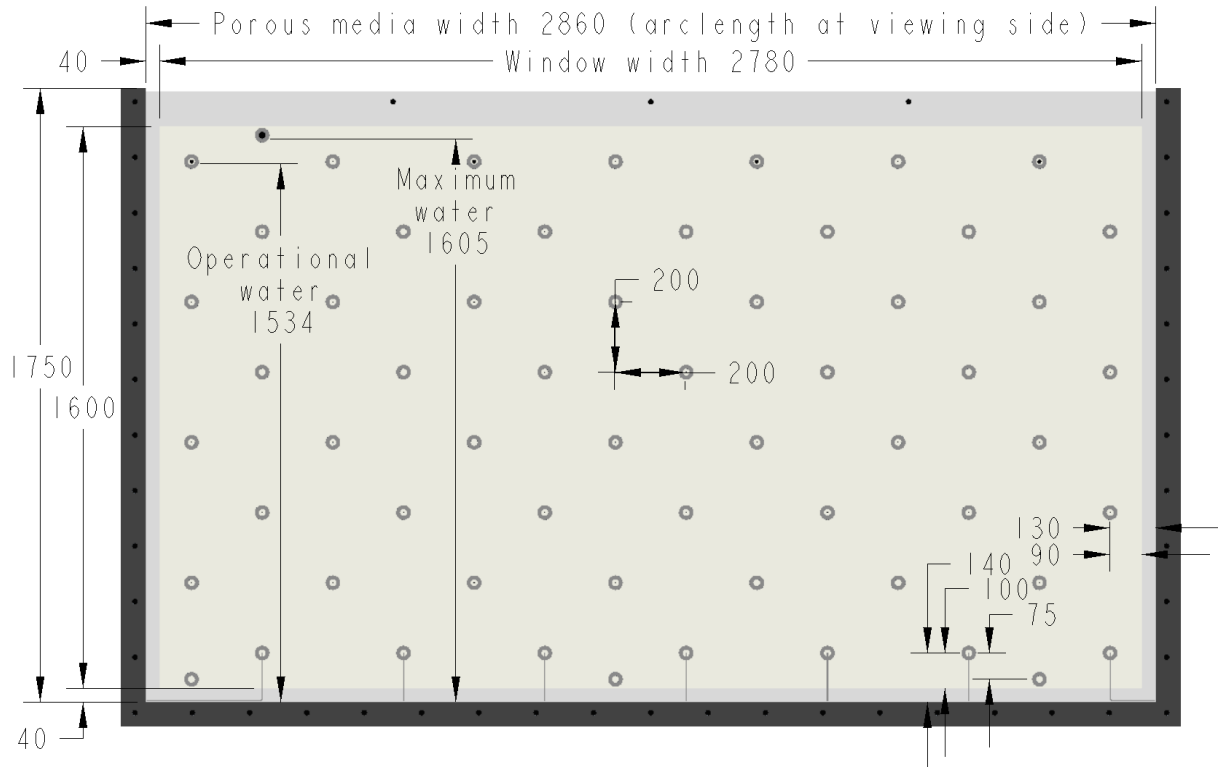

**Figure A3** – Perforations are located at every second intersection of a 200 · 200 mm square grid along the view side radius of the porous media (arclength at r = 3600 mm), translating to 201 · 200 mm (w · h) grid along the outer flow cell radius (r = 3618 mm), symmetrically distanced from the window boundaries. Three emergency drains are located at the bottom and at there is an always-open port to avoid overfilling. The seven perforations in the lower row have their inner tube extended to the lower flow cell boundary

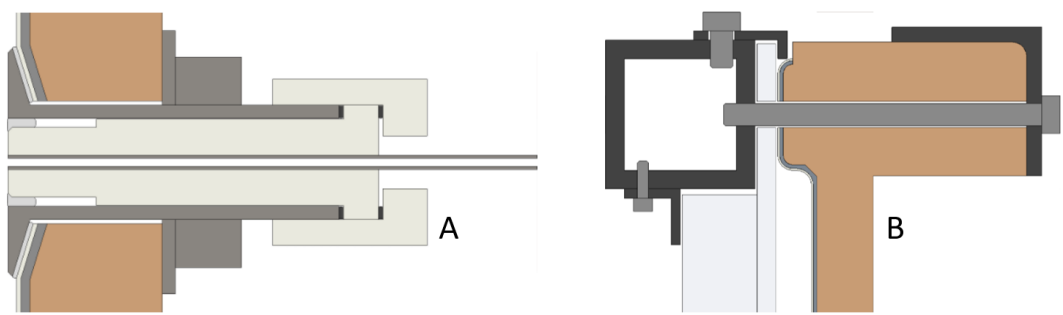

**Figure A4** - Cross sections of: A) Perforation as configured for sensor wells, whereas for injection wells the inner steel tube extends 10 mm into the system. The seven perforations in the lower row have their inner tube extended to the lower flow cell boundary. B) Frame cross section. 180 mm long M12 bolts connect rear and front frames. The inner sacrificial polycarbonate glass is clamped in place and the structural acrylic glass rests on a ledge inside the front frame. A in 2:1 scale relative to B. Both oriented with flow cell front towards left.

A4 – Forecasting study laboratory layout

The $CO_2$ experiments were performed in a ground-floor exhibit room within a semi-triangular plywood box (termed 'blackbox') to limit temperature variations (see **Figure A5**). The reflected section of the floor and ceiling, the personnel access point, and the wall immediately behind the cameras were all covered with black tapestry. The camera was positioned in the focal point of the curved viewing window, and high-frequency LED lamps were located at each lateral side of the flow cell at half its height. A levelling laser was used to produce the laser grid used for lens correction for image post processing. A control room was set up in an L-shape between the diagonal side of the box and the windows, with all windows blocked to obey to the double-blind criteria for the study. Multiple monitors were used for operation, and critical equipment was operated by batteries in case of power outages. The control room included fluid supply systems, with a water distribution manifold connected to the flush ports. A secondary camera provided a live view of the experiment.

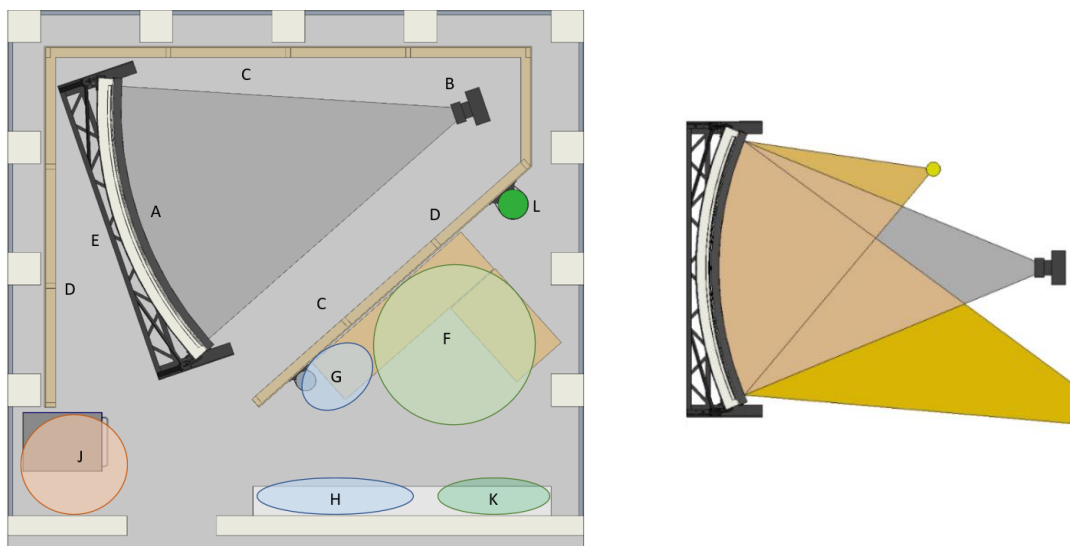

**Figure A5** – Left: Forecasting study laboratory layout and location of flow cell within the 'blackbox', where images were captured through the front window of the A) porous media geometry by the B) camera. C) Lamps were placed in front of the flow cell and temperature was regulated by D) panel ovens. Sensors and plumbing were connected to perforations on the E) rear side of the flow cell with. Fluid cycling and experiments were operated from the F) control room dry zone with assorted computers and monitors beside the F) fluid supply systems. Liquids were stored and samples were measured in the H) wet zone. Equipment repair and modifications were confined to the J) garage. K) Vacuum pump and air compressor location. L) Pressurized air for pneumatic equipment. Right: The curvature of room-scale version allows imaging with minimum reflection. Incident camera field of view overlaps horizontally with reflected field of view, and objects not directly between the camera and experiment remain hidden to the camera.

A5 – Forecasting study pH sensitive solution preparation protocol

An alkaline combination of sodium salts of BTB and MRe was used as the pH sensitive solution termed "FC green" with mixing proportions shown in **Table A3**. To minimize contamination and degradation, the solution were prepared using freshly deionized water in batches of cans at most 48 hours prior to injection (same requirement for the cleaning solution used for resetting the experiment). The FC green was prepared by dissolving and thoroughly mixing 0.80 g NaOH pellets into one liter of water. 200 mL of NaOH solution was watered to 2000 mL before adding 3.88 g of NaBTB and 5.24 g of NaMRe. The mixtures were magnet stirred and repeatedly shaken in closed containers before added to DI water (1000 mL added to 20 L). The mixture was then sealed, shaken vigorously and heated to approximately 30 °C to avoid gas release inside the flow cell. In solution the components dissociate to the compounds listed in **Table A4**. A fraction of the sodium ions would combine with $CO_2$ from air exposure, and the pH was therefore measured immediately prior to injection. After the $CO_2$ injection, 0.5 mM NaOH cleaning solution was prepared in batches by dissolving and thoroughly mixing 2.00 g NaOH into 1 L of water before adding 200 mL NaOH to cans containing approximately 20 L before adding water up to the 21 L mark.

**Table A3** - pH sensitive solution used in the $CO_2$ injection experiments.

| Salt compound | NaOH | NaBTB | NaMRe |
|---|---|---|---|
| Concentration $[mg \cdot L^{-1}]$ | 3.8 | 92 | 125 |
| Concentration $[\mu M]$ | 95 | 143 | 428 |
| Molar mass $[g \cdot mol^{-1}]$ | 40.00 | 646.36 | 291.28 |

**Table A4** – Composition of pH sensitive solution when injected. Ionic charge in superscript.

| Dissolved compound | $OH^-$ | $BTB^-$ | $MRe^-$ | $Na^+$ | $HNaCO_3$ |
|---|---|---|---|---|---|
| Concentration $[\mu M]$ | < 95 | 143 | 428 | 571-666 | < 95 |

A6 – Forecasting study experiment cycling

Fluid displacement is performed in a flushing sequence where the displacement front radiating from an active flush port is allowed to pass the next flush port with pre-determined margin to reduce the diluting effect of mixing zones. Ideally, this process should be standardized and scripted for the studied geometry to maintain chemical composition between experiments. A complete fluid cycling protocol is shown in **Figure A6** and detailed in **Table A5** for all $CO_2$ injections reported in Fernø et al., this issue.

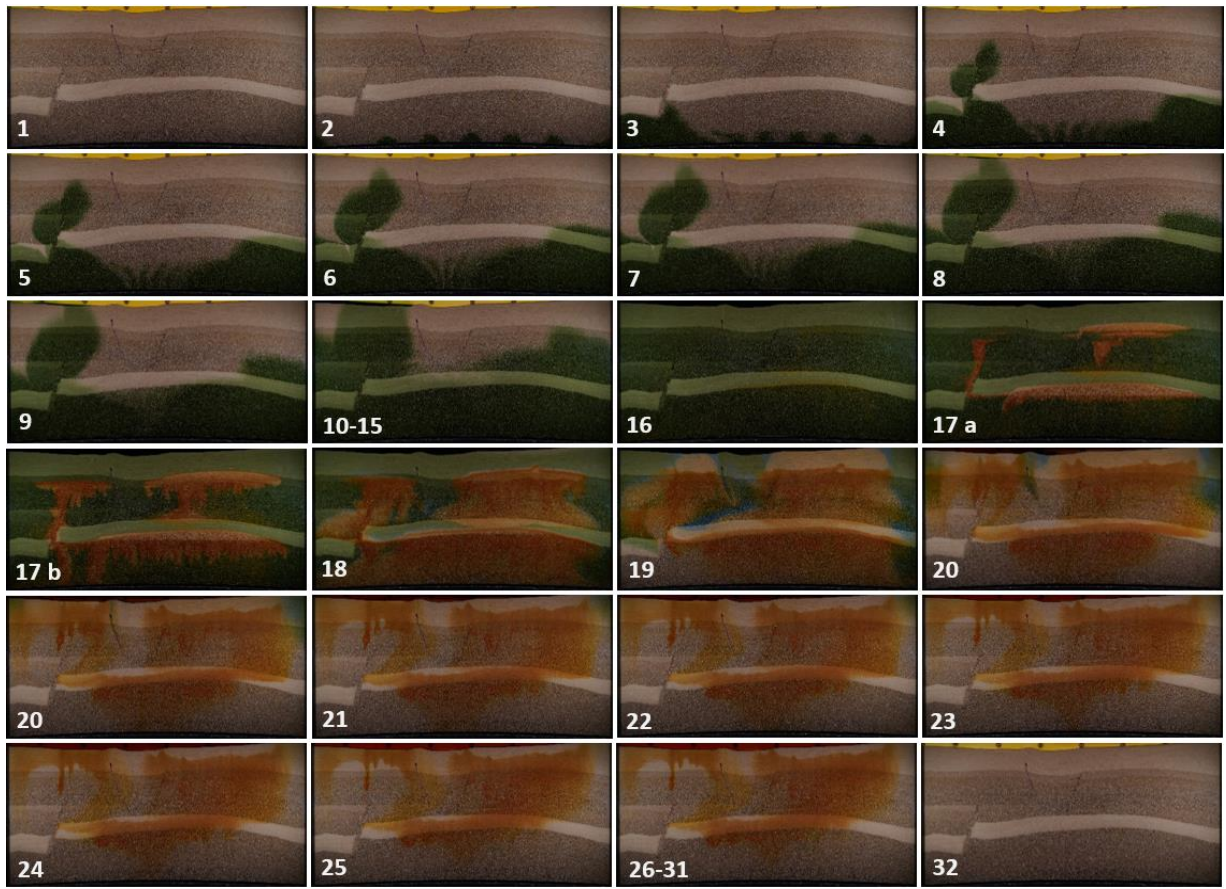

**Figure A6** – Experiment cycling from clean system before (image 1) and after (image 32) for experiment C5. Images 1 - 16 shows the injection of pH sensitive solution. Images 17 a and b shows different times during and after $CO_2$ injection. Images 18-32 display the cleaning process. Image numbers are described in **Table A5**.

**Table A5** - Fluid cycling in during CO2 injection, detailed with active flush ports and injected volumes.

| State | Flush ports [N, 7] active | | | | | | | Injection volumes [mL] | | | | | Step |
|---|---|---|---|---|---|---|---|---|---|---|---|---|---|
| Clean | *Sequence initiation date* | | | | | | | *Oct 27* | *Dec 02* | *Dec 13* | *Dec 23* | *Jan 03* | 1 |
| Injection of FC green solution | 1 | 3 | 5 | 7 | 9 | 11 | 13 | | 1000 | 1000 | 1000 | 1000 | 2 |
| | 1 | | | | | | | | 8000 | 2000 | 2000 | 2000 | 3 |
| | 1 | | | | | | 13 | | 19000 | 6000 | 6000 | 6000 | 4 |
| | | 3 | | | | 11 | | | 9500 | 2000 | 2000 | 2000 | 5 |
| | | 3 | | | 9 | | | | 9500 | 3000 | 3000 | 3000 | 6 |
| | | 3 | | | | | | | 2000 | 1000 | 1000 | 1000 | 7 |
| | | | 5 | | 9 | | | | 7500 | 3000 | 3000 | 3000 | 8 |
| | | | | | 9 | | | n. a. [b] | 875 | 875 | 875 | 875 | 9 |
| | 1 | | 5 | 7 | | | | | 2000 | 2000 | 2000 | 2000 | 10 |
| | 1 | | 5 | | | | | | 875 | [e] 1610 | 875 | 875 | 11 |
| | 1 | | 5 | 7 | | | | | 2000 | 2000 | 2000 | 2000 | 12 |
| | 1 | | 5 | | | | | | 875 | 875 | 875 | 875 | 13 |
| | 1 | | 5 | 7 | | | | | 2000 | 2000 | 2000 | 2000 | 14 |
| | 1 | | 5 | | | | | | 875 | 875 | 875 | 875 | 15 |
| | | | | 7 | | | | | 38000 | 41890 | 31611 | 28500 | 16 |
| Green | *Total injected volume* | | | | | | | 353000 | 104000 | 71000 | 59111 | 56000 | *n. a.* |
| | *Sequence completion date* | | | | | | | *Nov 23* | *Dec 04* | *Dec 14* | *Dec 24* | *Jan 04* | *n. a.* |
| Experiment | *FC cycle no.* [a] | | | | | | | **C 1** | **C 2** | **C 3** | **C 4** | **C 5** | 17 |
| Injection of cleaning solution | *Date* | | | | | | | *Nov 29* | *Dec 09* | *Dec 20* | *Dec 30* | *Jan 09* | *n. a.* |
| | 1 | 3 | 5 | 7 | 9 | 11 | 13 | 1000 | 1000 | 1000 | 1000 | 1000 | 18 |
| | 1 | | | | | | | 8000 | 8000 | 8000 | 8000 | 8000 | 19 |
| | 1 | | | | | | 13 | 9500 | 19000 | 19000 | 19000 | 19000 | 20 |
| | | 3 | | | | 11 | | [c] 19000 | 9500 | 9500 | 9500 | 9500 | 21 |
| | | 3 | | | 9 | | | [d] 19000 | 9500 | 9500 | 7500 | 7500 | 22 |
| | | 3 | | | | | | 2000 | 2000 | 2000 | 2000 | 2000 | 23 |
| | | | 5 | | 9 | | | 7500 | 7500 | 7500 | 7500 | 7500 | 24 |
| | | | | | 9 | | | 875 | 875 | 875 | 875 | 875 | 25 |
| | 1 | | 5 | 7 | | | | 1750 | 2875 | 2000 | 2000 | 2000 | 26 |
| | 1 | | 5 | | | | | 875 | 875 | 875 | 875 | 875 | 27 |
| | 1 | | 5 | 7 | | | | 2292 | 2000 | 2000 | 2000 | 2000 | 28 |
| | 1 | | 5 | | | | | 875 | 875 | 875 | 875 | 875 | 29 |
| | 1 | | 5 | 7 | | | | 2292 | 2000 | 2000 | 2000 | 2000 | 30 |
| | 1 | | 5 | | | | | 875 | 875 | 875 | 875 | 875 | 31 |
| | | | | 7 | | | | 133000 | 209000 | 209000 | 209000 | 209000 | 32 |
| Clean system | *Total injected volume* | | | | | | | 208834 | 275875 | 275000 | 273000 | 273000 | *n. a.* |
| | *Sequence completion date* | | | | | | | *Dec 02* | *Dec 13* | *Dec 23* | *Jan 03* | *Jan 13* | *n. a.* |

[a] Detailed in (Fernø et al., this issue).
[b] Sequence significantly deviating from remainder, with FC green injected a second time after it became apparent that the first injected FC green had been significantly degraded by air exposure and water source contamination. Uncertainties described in [Fernø]. Estimated total volume.
[c] In ports [1, 7], [3, 7], [11, 7], and [13, 7].
[d] In ports [1, 7], [3, 7], [9, 7], and [11, 7].
[e] Valve not opened for next step. Correction injection of 875 mL through port [7, 9], additional step not listed in table.
Last column (Jan 03 - 13, cycle C5) corresponds to images shown in Figure A6

## A7 – Supplementary forecasting study data

Water phase samples were collected from ports marked in **Figure A7** immediately prior to gas injections, with pH values listed in **Table A6**. Overflow logging was implemented for experiment cycles C3, C4, and C5, producing results such as that shown in **Figure A7**. After the Forecasting study experiment series, another series of well tests were performed using FC green as tracer, presented in **Figure A8**. Prior to flow cell draining, laboratory dismantling, and depth measurements (method shown in **Figure A8**), a series of buffer solutions were injected to provide reference colors for image post-processing, shown in **Figure A9**.

**Table A7** - Measured pH of water phase sampled from ports prior to Forecasting study experiments. Uncertainty ± 0.2.

| $[x,y]$ | | 0, 4 | 0, 6 | 1, 5 | 2, 6 | 5, 3 | 6, 4 | 6, 6 | 7, 3 | 9, 3 | 9, 5 | 10, 4 | 12, 4 | 12, 6 |
|---|---|---|---|---|---|---|---|---|---|---|---|---|---|---|
| C2 | | 7.80 | - | 7.87 | 7.84 | 7.95 | 8.29 | 8.10 | 8.10 | 7.91 | 7.80 | - | - | 7.71 |
| C4 | | 9.61 | - | 8.65 | - | 10.22 | - | 9.01 | 10.12 | 9.8 | - | - | *9.45* | 8.72 |
| C5 | 1st | 9.79 | 8.71 | | | 10.20 | | 8.84 | 10.09 | 9.74 | 8.72 | 9.47 | 9.58 | 8.80 |
| | 2nd | 9.83 | 9.02 | | | 10.22 | | 8.86 | 10.04 | 9.82 | 8.69 | 9.53 | 9.62 | 8.62 |

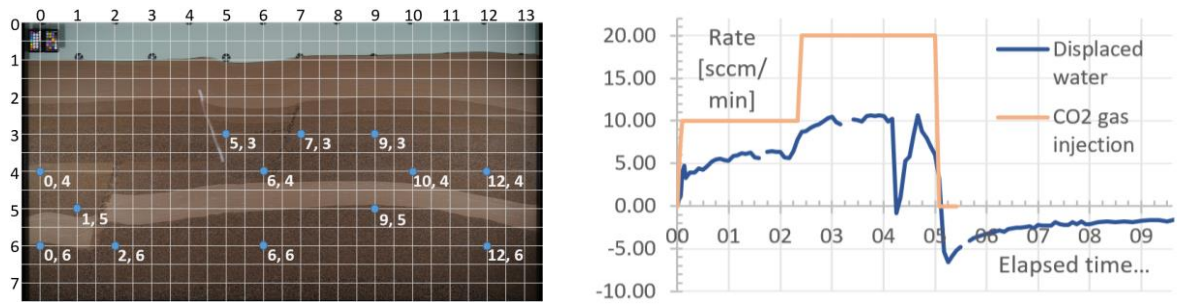

**Figure A7** - Water phase sampling ports (left) listed in **Table A6**. Overflow rate for cycle C3 (right).

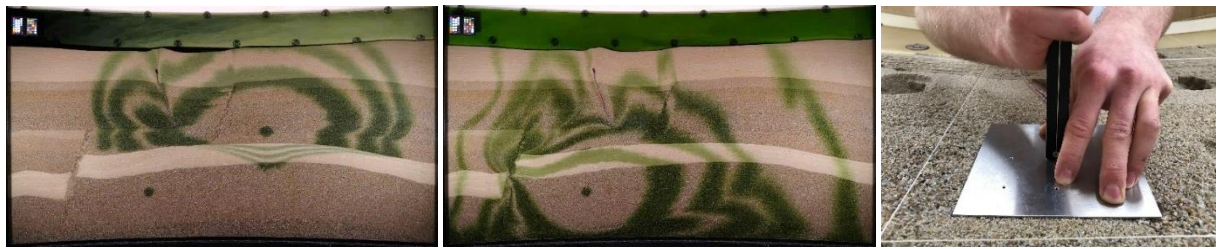

**Figure A8** – Left and center: Forecasting study geometry well tests after conclusion of main experiment series using FC green as tracer in clean water. Composite images. Right: Point depth measurements were collected (center) while sands were still moist to produce the depth map presented in (Fernø et al., this issue).

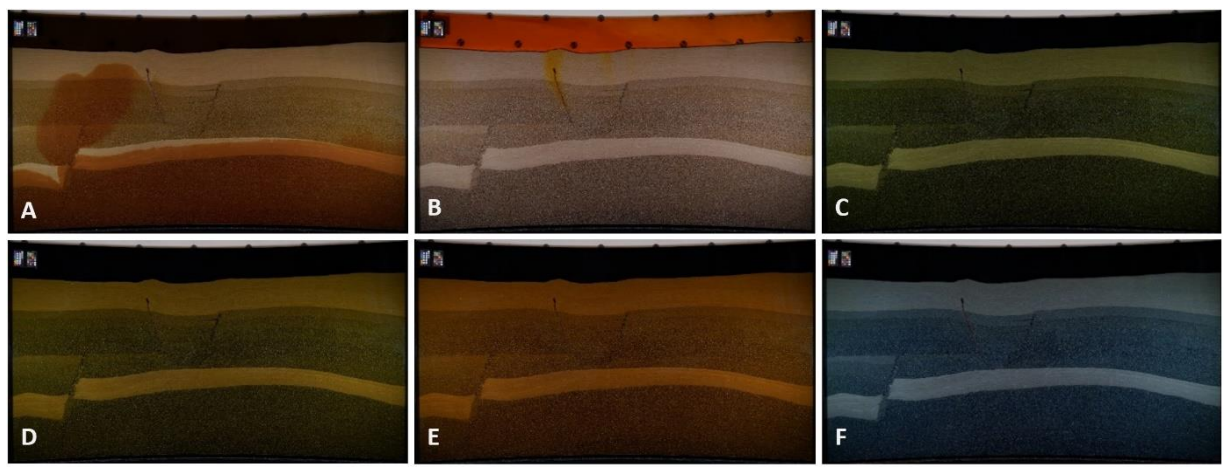

**Figure A9** – Buffer solutions injected into the Forecasting study geometry after conclusion of the Forecasting experiment series. Injection protocols similar to those presented in **table A5**, and BTB/MRe concentrations as in FC green. A) pH 5.84-5.85, 65 L injected with phase separation observed. B) Without both BTB and MRe, pH 7.89-8.06, 72 L injected. C) pH 7.96-7.98, 72 L injected. D) pH 6.99-7.1, 103 L injected. E) pH 6.45-6.49, 99 L injected. F) Without MRe, pH 8.00-8.04.

**Appendix B – Tabletop FluidFlower**

This appendix details the tabletop FluidFlower. Different versions exist and have been used for rapid prototyping and methodology development, as well as experiment series such as those detailed in (Haugen et al, this issue; Saló-Salgado et al., this issue, Keilegavlen et al, this issue). Specifications presented apply to the second such flow cell built (in-house designation FF3.2 "Bilbo").

B1 – Dimensions of tabletop FluidFlower

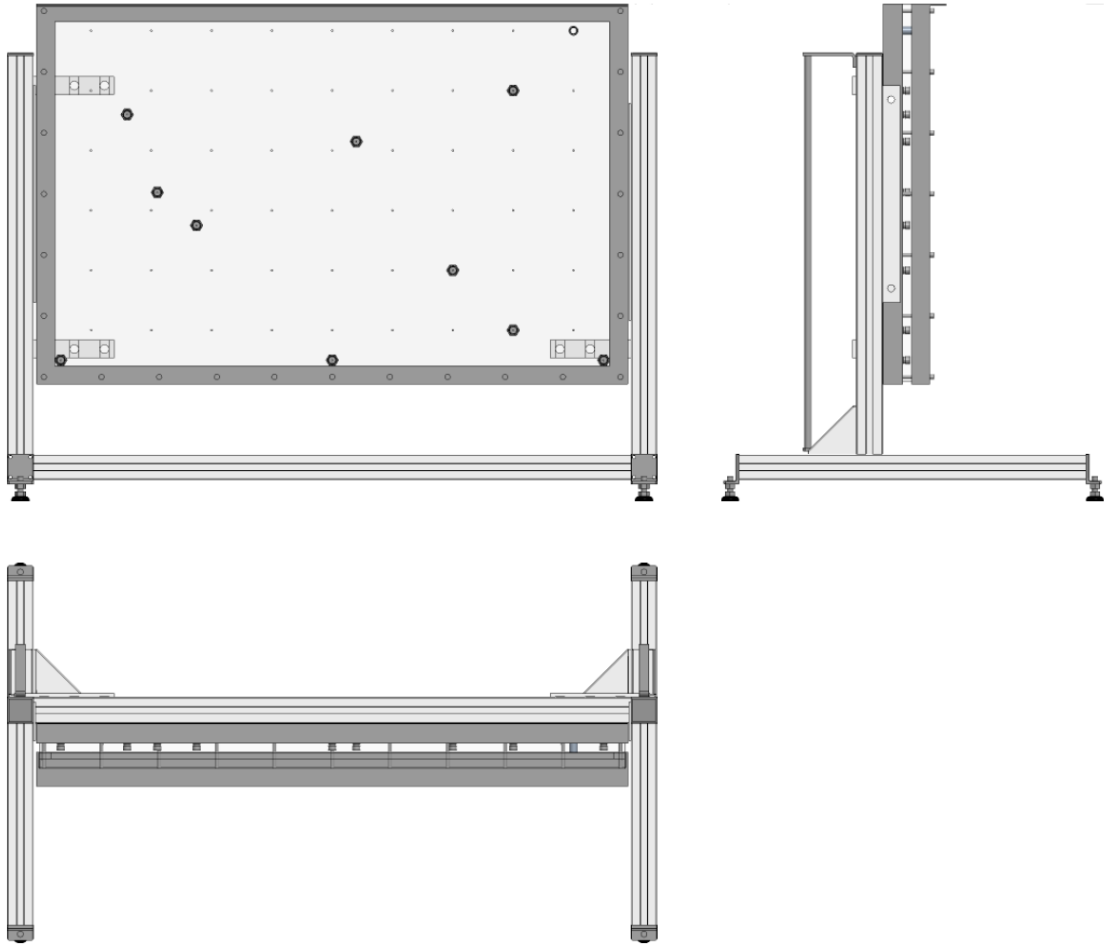

**Figure B1** – Front, side, and top profile of tabletop FluidFlower. Instrumentation rails are attached to the substructure.

**Table B1** – Dimensions of tabletop FluidFlower.

| | | |
|---|---|---|
| **Internal measurements** | Height, total [mm] | 612 |
| | Height of water table [mm] | 563 |
| | Width [mm] | 934 |
| | [a]Depth [mm] (empty) | 10 |
| | Volume (empty) under operational water level [L] | 5.3 |
| | Volume (empty), total [L] | 5.7 |
| **Window measurements** | Height, total [mm] | 575 |
| | Height of visible water table [mm] | 556 |
| | Width [mm] | 920 |
| | Area, total [m$^2$] | 0.53 |
| **Full structure measurements** | Height [mm] (minimum configuration) | 709 |
| | Width [mm] | 1078 |
| | Depth [mm] (without legs) | 575 |
| | Mass, empty [kg] (approximate) | 30 |
| [a] Only accurate when the flow cells are not under strain (filled with liquids and/or sand) | | |

B2 – Material choices for tabletop FluidFlower

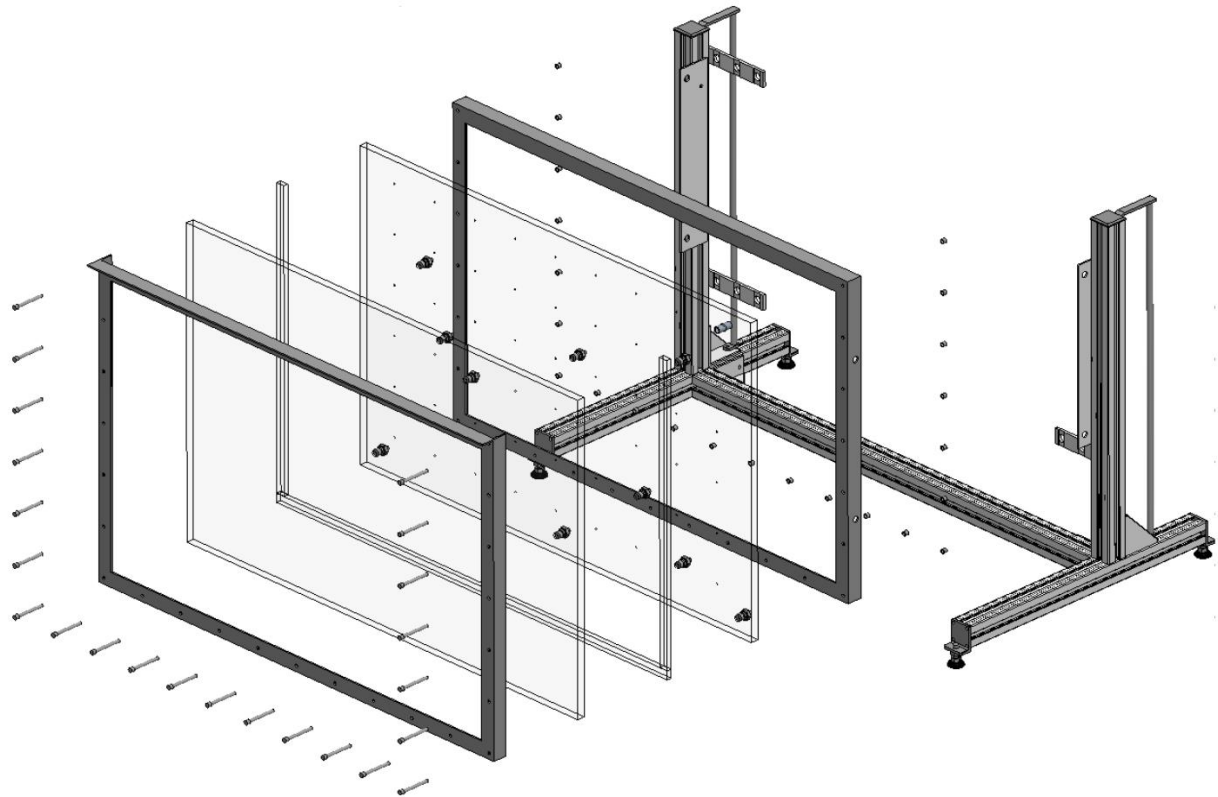

**Figure B2** – Exploded assembly of flow cell components. Acrylic windows with spacers between them are clamped between the steel frames. Bolts and nuts clamp the circumference of the flow cell. The rear frame connects to the FlexLink substructure, onto which valve manifolds and instrumentation rails are connected.

**Table B2** – Materials in tabletop FluidFlower.

| **Internal materials** | Front wall | Acrylic glass |
|---|---|---|
| | Rear wall | Acrylic glass |
| | Bottom/sides | Acrylic glass |
| | Rear corners | Polyurethane sealant |
| | Front corners | Polyurethane sealant |
| | Perforation seal | External o-ring |
| | Perforations | Stainless steel |
| **External materials** | Flange frames | Stainless steel |
| | Instrumentation rails | Stainless steel rods |
| | Substructure | Aluminium |

B3 – Perforations in tabletop FluidFlower

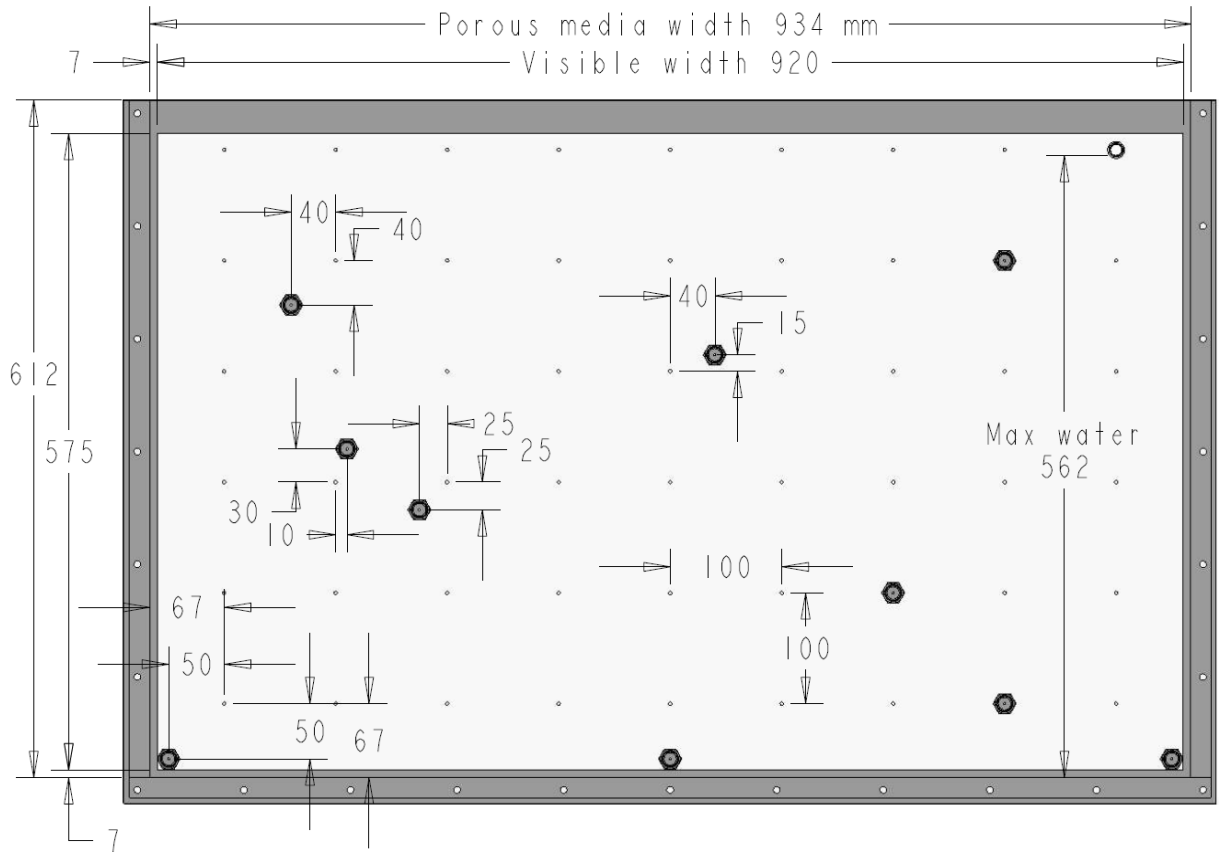

**Figure B3 –** Perforations in the tabletop FluidFlowers as configured for the PoroTwin AI controlled experiment series. Additional perforations are comparatively easily added between experiment series, and therefore these versions have been perforated on an as-needed basis.

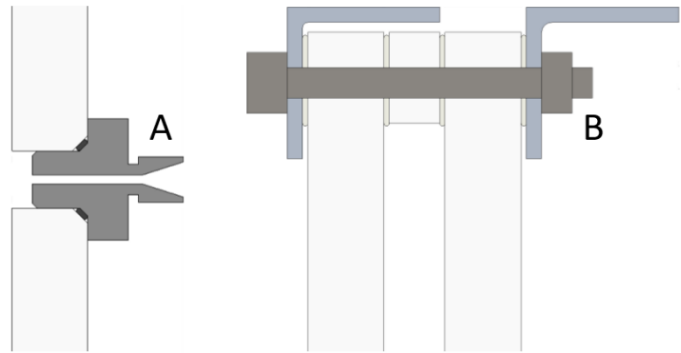

**Figure B4** – A) Perforation cross section as configured for Forecasting study and legacy experiments. Perforations are external-gasket standard stainless steel double male parallel NPT ½'' to 1/8'' Swagelok connectors, for post-Forecasting study configurations with a wedge lodged into the internal volume from the front-facing side so that sands are not removed from the system during fluid suction. B) Frame cross section. 60 mm long M6 bolts with nuts secure all layers in the structure.

B4 – Experiment cycling in tabletop FluidFlower

Fluid displacement is performed in a flushing sequence following the same principle as that shown in Appendix A. The smaller scale of the tabletop FluidFlower allows for a simplified sequence, and the introduction of wedged perforations reduces the risks associated with fluid production. The displacement front radiating from an active flush port is allowed to pass the next flush port to reduce the diluting effect of mixing zones. If this is not accounted for, local differences in the water-phase chemistry may become non-negligible and reduce the physical reproducibility of observed flow. Multiple perforations are used for sequential injection to reduce the required cycled fluid volume, producing minimal waste while promoting fluid homogeneity.

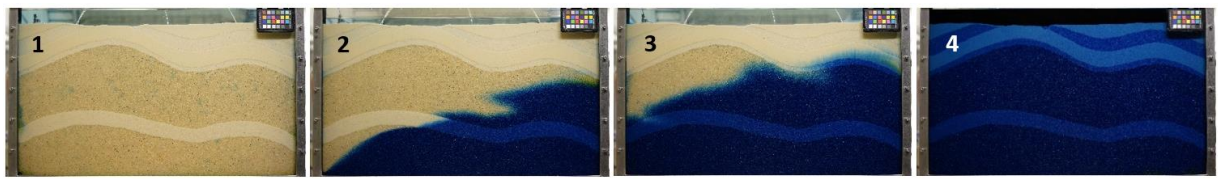

**Figure B5** – Example displacement of cleaning solution by BTB solution: 1) System cleaned post-experiment. Simultaneous injection in all ports to clean injection tubes and manifold is followed by injection right port with production in middle, then left, and after a while injection is moved to middle port. When 2) the lower porous media is fully saturated, production is stopped to allow displacement front to pass leftmost port before 3) injection is moved to left port (notice mixing zone radiating from bottom left corner). Injection continues until 4) system is saturated.

**Table B3** – Example repeated multiphase experiment cycling checklist.

| Typical time | Step | |
|---|---|---|
| Day before experiment | 1 | Prepare water solution to experiment specification. |
| | 2 | Sample pH of prepared solution. |
| | 3 | Degas water solution to experiment specification. |
| | 4 | Rinse pump with solution to be injected. |
| | 5 | Empty, rinse, refill, and bleed gas from gas trap. |
| | 6 | Update flush protocol to current time. |
| | 7 | Initiate interval imaging of flush sequence. Synchronize interval to protocol. |
| | 8 | Inject solution according to experiment series protocol. |
| Experiment day | 9 | Prepare water supply system for overflow and logging. |
| | 10 | Open gas supply and control that pressure is within acceptable range. |
| | 11 | Direct gas flow to mass flow controller. Pre-warm MFC by directing gas flow to room. |
| | 12 | Initiate pressure and temperature logging. |
| | 13 | Rinse injection and sampling ports. |
| | 14 | Collect and measure pH samples. |
| | 15 | Inject water in free water phase until active overflow. |
| | 16 | Initiate scripted overflow and mass logging when no active overflow. |
| | 17 | Check that mass flow controller performs nominally. Perform dummy ramp. |
| | 18 | Initiate mass flow controller logging. |
| | 19 | Synchronize offline clocks. |
| | 20 | Update experiment protocol to current time. |
| | 21 | Capture high-resolution image. |
| | 22 | Initiate interval imaging of experiment. Synchronize interval to protocol. |
| | 23 | Connect gas supply to inlet and pressurize gas line to system hydrostatic pressure. |
| | 24 | Perform experiment according to protocol. Confirm steps. |
| | 25 | Close and disconnect gas supply from inlet. |
| | 26 | Change imaging interval for resting phase of experiment. |
| Days after | 27 | End instrument logging when redundant. |
| | 28 | Prepare and degas cleaning solution. |
| | 29 | Rinse pump with cleaning solution. |
| | 30 | Empty, rinse, refill, and bleed gas from gas trap. |
| | 31 | Update flush protocol to current time. |
| | 32 | Initiate interval imaging of flush sequence. Synchronize interval to protocol. |
| | 33 | Inject solution according to experiment series protocol. |